\documentclass{ws-ijmpps}

\begin{document}

\markboth{BO-QIANG MA} {Neutrino Oscillations: from Standard and
Non-standard Viewpoints}

\catchline{}{}{}{}{}

\title{Neutrino Oscillations: from Standard and Non-standard Viewpoints}

\author{\footnotesize BO-QIANG MA}

\address{School of Physics and State Key Laboratory of Nuclear Physics and
Technology, Peking~University, Beijing 100871, China\\
mabq@pku.edu.cn}

\maketitle


\begin{abstract}
In the standard model of neutrino oscillations, the neutrino flavor
states are mixtures of mass-eigenstates, and the phenomena are well
described by the neutrino mixing matrix, i.e., the PMNS matrix. I
review the recent progress on parametrization of the neutrino mixing
matrix. Besides that I also discuss on the possibility to describe
the neutrino oscillations by a non-standard model in which the
neutrino mixing is caused by the Lorentz violation (LV) contribution
in the effective field theory for LV. We assume that neutrinos are
massless and that neutrino flavor states are mixing states of energy
eigenstates. In our calculation the neutrino mixing parts depend on
LV parameters and neutrino energy. The oscillation amplitude varies
with the neutrino energy, thus neutrino experiments with energy
dependence may test and constrain the Lorentz violation scenario for
neutrino oscillation.

\end{abstract}

\ccode{PACS Nos.: 14.60.Pq, 11.30.Cp, 12.15.Ff, 14.60.Lm}

\vspace{6mm}


The mixing between different generations of quarks and leptons is
one of the most fundamental and important issues in particle
physics. Parametrization of mixing matrices is an important step to
understand the mixing of fermions. The Cabibbo\cite{cabibbo},
Kobayashi and Maskawa\cite{km}(CKM) matrix $V_{\mathrm{CKM}}$
describes the mixing of quarks of three generations, and the mixing
could be regarded as a rotation from fermion mass eigenstates to
flavor eigenstates. The abundant experimental data on neutrinos
convincingly suggest the mixing of different generations of
neutrinos, just analogous to that of quarks. To explain neutrino
oscillations, the conventional scenario is to assume that neutrinos
have masses. From this assumption, there is a spectrum of three or
more neutrino mass eigenstates and the flavor state is the mixing
state of mass eigenstates, and lepton mixing is described by the
Pontecorvo\cite{pontecorvo}-Maki-Nakawaga-Sakata\cite{mns} (PMNS)
matrix $U_{\rm{PMNS}}$.

A commonly used form of mixing matrix for three generations of
fermions is given by~\cite{ck,yao},
\begin{eqnarray}
V(\mathrm{or}~U) = \left(
\begin{array}{ccc}
c_{12}c_{13} & s_{12}c_{13} & s_{13}e^{-i\delta}           \\
-s_{12}c_{23}-c_{12}s_{23}s_{13}e^{i\delta} &
c_{12}c_{23}-s_{12}s_{23}s_{13}e^{i\delta}  & s_{23}c_{13} \\
s_{12}s_{23}-c_{12}c_{23}s_{13}e^{i\delta}  &
-c_{12}s_{23}-s_{12}c_{23}s_{13}e^{i\delta} & c_{23}c_{13}
\end{array}
\right),\label{fp}
\end{eqnarray}
where $s_{ij}=\sin\theta_{ij}$ and $c_{ij}=\cos\theta_{ij}$ are the
mixing angles and $\delta$ is the CP violating phase. If neutrinos
are of Majorana type, for the PMNS matrix one should include an
additional diagonal matrix with two Majorana phases ${\rm
diag}(e^{i\alpha_1/2},e^{i\alpha_2/2},1)$ multiplied to the matrix
from right in the above. The CKM matrix is close to the unit matrix
with small deviations in the non-diagonal elements. This led
Wolfenstein to parameterize the quark mixing matrix
as~\cite{wolfensteinpara}, {\small
\begin{eqnarray} V_{\mathrm{CKM}}=\left(
  \begin{array}{ccc}
    1-\frac{1}{2}\lambda^2   & \lambda                & A\lambda^3(\rho-i\eta) \\
    -\lambda                 & 1-\frac{1}{2}\lambda^2 & A\lambda^2             \\
    A\lambda^3(1-\rho-i\eta) & -A\lambda^2            & 1                      \\
  \end{array}
\right)+\mathcal{O}(\lambda^4).\label{wolfenstein}
\end{eqnarray}
This parametrization explicitly shows the deviations of the
non-diagonal elements from the unit matrix in different powers of
the parameter $\lambda$ with $\lambda=0.2257^{+0.0009}_{-0.0010}$.
The other parameters are $A=0.814^{+0.021}_{-0.022}$,
$\rho(1-\lambda^2/2+\dots) = 0.135^{+0.031}_{-0.016}$, and
$\eta(1-\lambda^2/2+\dots) = 0.349^{+0.015}_{-0.017}$~\cite{yao}.
The parameter $\lambda$ serves as a good indicator of hierarchy of
the mixing phenomenon in quark sector.

Quite different from quark mixing matrix, almost all the
non-diagonal elements of the neutrino mixing matrix are large, only
with the exception of  $V_{e3}$. So it is unpractical to expand the
matrix in powers of one of the non-diagonal elements, like the
Wolfenstein parametrization of the quark mixing matrix. In practice
one may parameterize the neutrino mixing matrix with other bases,
such as
\begin{eqnarray}
    \left(
        \begin{array}{ccc}
            \sqrt{2}/2 & \sqrt{2}/2 & 0 \\
            -1/2 & 1/2 & \sqrt{2}/2 \\
            1/2 & -1/2 & \sqrt{2}/2
        \end{array}
        \right),
    \left(
        \begin{array}{ccc}
            \sqrt{6}/3 & \sqrt{3}/3 & 0 \\
            -\sqrt{6}/6 & \sqrt{3}/3 & \sqrt{2}/2 \\
            \sqrt{6}/6 & -\sqrt{3}/3 & \sqrt{2}/2
        \end{array}
        \right),\nonumber
\end{eqnarray}
which are called the bimaximal mixing pattern and the tri-bimaximal
pattern respectively. There have been many different forms of
parametrization\cite{rodejohann,linan}, and finding one which is
simple and convenient to use is important as a tool for further
theoretical and experimental studies.

There have been progress to parameterize the lepton mixing matrix
based on idea of connection between the quark and lepton mixing
matrices. The quark-lepton complementary (QLC) is an interesting
example in this direction. The quark-lepton complementarity
(QLC)\cite{smirnov,qlc} relates quark and lepton mixing angles with
\begin{eqnarray}
&&\theta_{12}^Q+\theta_{12}^L=\frac{\pi}{4},\nonumber\\
&&\theta_{23}^Q+\theta_{23}^L=\frac{\pi}{4},\nonumber\\
&&\theta_{13}^Q\sim\theta_{13}^L\sim0,\label{qlc}
\end{eqnarray}
where the superscript $Q$ indicates the mixing angles in the CKM
matrix, and the superscript $L$ indicates mixing angles in the PMNS
mixing matrix $U_{\mathrm{PMNS}}$. The above relations enable one to
express the mixing parameters of lepton mixing based on information
of quark mixing parameters.

Nan Li and I\cite{LIN} performed the parametrization of the PMNS
matrix based on this idea and got the following results in two cases
by combing QLC with the Wolfenstein parametrization of CKM matrix.
For Case 1, i.e.,
$\sin\theta_2^{\mathrm{PMNS}}e^{-i\delta}=A\lambda^3(\zeta-i\xi)$,
we get the PMNS matrix
\begin{eqnarray}\nonumber
   U&=&
\left(
        \begin{array}{ccc}
            \frac{\sqrt{2}}{2} & \frac{\sqrt{2}}{2} & 0 \\
            -\frac{1}{2} & \frac{1}{2} & \frac{\sqrt{2}}{2} \\
            \frac{1}{2} & -\frac{1}{2} & \frac{\sqrt{2}}{2}
        \end{array} \right)+\lambda
\left(
        \begin{array}{ccc}
             \frac{\sqrt{2}}{2} & -\frac{\sqrt{2}}{2} & 0 \\
            \frac{1}{2} & \frac{1}{2} &  0\\
           - \frac{1}{2} & -\frac{1}{2} & 0
        \end{array} \right)+\lambda^{2}
\left(
        \begin{array}{ccc}
            -\frac{\sqrt{2}}{4} & -\frac{\sqrt{2}}{4} & 0 \\
            -\frac{1}{2}(A-\frac{1}{2}) & \frac{1}{2}(A-\frac{1}{2}) & -\frac{\sqrt{2}}{2}A \\
            -\frac{1}{2}(A+\frac{1}{2}) & \frac{1}{2}(A+\frac{1}{2}) & \frac{\sqrt{2}}{2}A
        \end{array} \right)\nonumber\\&&+\lambda^{3}
\left(
        \begin{array}{ccc}
            0 & 0 & A(\zeta-i\xi) \\
            \frac{1}{2}A(1-\zeta-i\xi) & \frac{1}{2}A(1-\zeta-i\xi) & 0\\
            \frac{1}{2}A(1-\zeta-i\xi) & \frac{1}{2}A(1-\zeta-i\xi) & 0
        \end{array} \right)+\mathcal{O}(\lambda^4).\label{eq.3ci}
\end{eqnarray}
For Case 2, i.e.,
$\sin\theta_2^{\mathrm{PMNS}}e^{-i\delta}=A\lambda^2(\zeta-i\xi)$,
we get
\begin{eqnarray}\nonumber
   U&=&
\left(
        \begin{array}{ccc}
            \frac{\sqrt{2}}{2} & \frac{\sqrt{2}}{2} & 0 \\
            -\frac{1}{2} & \frac{1}{2} & \frac{\sqrt{2}}{2} \\
            \frac{1}{2} & -\frac{1}{2} & \frac{\sqrt{2}}{2}
        \end{array} \right)+\lambda
 \left(
        \begin{array}{ccc}
             \frac{\sqrt{2}}{2} & -\frac{\sqrt{2}}{2} & 0 \\
            \frac{1}{2} & \frac{1}{2} & 0\\
           - \frac{1}{2} & -\frac{1}{2} & 0
        \end{array} \right)\nonumber\\&&+\lambda^{2}
 \left(
        \begin{array}{ccc}
            -\frac{\sqrt{2}}{4} & -\frac{\sqrt{2}}{4} & A(\zeta-i\xi) \\
            \frac{1}{2}[\frac{1}{2}-A(1+\zeta+i\xi)] & -\frac{1}{2}[\frac{1}{2}-A(1-\zeta-i\xi)] & -\frac{\sqrt{2}}{2}A\\
            -\frac{1}{2}[\frac{1}{2}+A(1+\zeta+i\xi)] & \frac{1}{2}[\frac{1}{2}+A(1-\zeta-i\xi)] & \frac{\sqrt{2}}{2}A
        \end{array} \right)+\mathcal{O}(\lambda^3).\label{eq.2ci}
\end{eqnarray}

From which we have the following observations: (1). The bimaximal
mixing pattern is derived naturally as the leading-order
approximation. (2). The Wolfenstein parameter $\lambda$ can
characterize both the deviation of the CKM matrix from the unit
matrix, and the deviation of the PMNS matrix from the exactly
bimaximal mixing pattern. More explicitly, the range of $\lambda$ in
PMNS matrix is calculated: $0.11<\lambda^{\mathrm{PMNS}}<0.24 $. In
this unified parametrization, $\lambda$ here is just the Wolfenstein
parameter of the CKM matrix,
$\lambda^{\mathrm{CKM}}=\sin\theta^{\mathrm{CKM}}=0.2243$. We can
see that the above values are consistent with each other compared
with the experimental data. However, we can also take the parameter
$\lambda^{\mathrm{PMNS}}$ as being not the same as the Wolfenstein
parameter $\lambda^{\mathrm{CKM}}$, and the symmetry between the
quark and lepton mixing matrices will break slightly.

Current experimental data show that the $U_{\mathrm{PMNS}}$ matrix
is close to the tri-bimaximal pattern. Parametrization of the PMNS
matrix around the tri-bimaximal pattern leads to more fast
converging expansions~\cite{linan}. For an unified parametrization
of both lepton and quark mixing matrices, Shi-Wen Li and
I\cite{LISW} introduced another method of the parametrization of the
CKM matrix. A new matrix was introduced instead of the unit matrix
as the basis of the CKM matrix~\cite{LISW},
\begin{eqnarray}
    V_b=\left(
        \begin{array}{ccc}
            \frac{\sqrt{2}+1}{\sqrt{6}} & \frac{\sqrt{2}-1}{\sqrt{6}} & 0 \\
           -\frac{\sqrt{2}-1}{\sqrt{6}} & \frac{\sqrt{2}+1}{\sqrt{6}} & 0 \\
            0                           & 0                           & 1
        \end{array}
        \right). \label{vb}
\end{eqnarray}
Though this new matrix is a little more complicated than the unit
matrix, it is closer to reality. The deviations of the CKM matrix
from the new matrix are rather small and the expansion converges
very quickly. We have proved that this matrix can be combined wirh
QLC to arrive at new parametrization of PMNS matrix with bases of
tri-bimaximal pattern.

The new  parametrization of the CKM matrix reads
\begin{eqnarray}
    V&=&\left(
        \begin{array}{ccc}
            \frac{\sqrt{2}+1}{\sqrt{6}} & \frac{\sqrt{2}-1}{\sqrt{6}} & 0 \\
           -\frac{\sqrt{2}-1}{\sqrt{6}} & \frac{\sqrt{2}+1}{\sqrt{6}} & 0 \\
            0                           & 0                           & 1
        \end{array}
        \right)+\lambda\left(
        \begin{array}{ccc}
            -(3-2\sqrt{2})               & 1                             & 0 \\
            -1                           & -(3-2\sqrt{2})                & A \\
            \frac{\sqrt{2}-1}{\sqrt{6}}A & -\frac{\sqrt{2}+1}{\sqrt{6}}A & 0
        \end{array}\right)\nonumber\\
        &&+\lambda^2\left(
        \begin{array}{ccc}
            -(30\sqrt{3}-21\sqrt{6})        & 0                                 & (\rho-i\eta)A \\
            \frac{\sqrt{2}-1}{2\sqrt{6}}A^2 &-(30\sqrt{3}-21\sqrt{6})-\frac{\sqrt{2}+1}{2\sqrt{6}}A^2 & 0 \\
            \left(1-\frac{\sqrt{2}+1}{\sqrt{6}}(\rho+i\eta)\right)A
            & \left(3-2\sqrt{2}-\frac{\sqrt{2}-1}{\sqrt{6}}(\rho+i\eta)\right)A & -{1\over2}A^2
        \end{array}\right)+\mathcal{O}(\lambda^3).~~~~\label{ve}
\end{eqnarray}

For the PMNS matrix, we get the new parametrization with
tri-bimaximal pattern in two cases:

\begin{eqnarray}
U&=&\left(
  \begin{array}{ccc}
    \frac{\sqrt{6}}{3}  & \frac{\sqrt{3}}{3} & 0                  \\
    -\frac{\sqrt{6}}{6} & \frac{\sqrt{3}}{3} & \frac{\sqrt{2}}{2} \\
    \frac{\sqrt{6}}{6}  & -\frac{\sqrt{3}}{3}& \frac{\sqrt{2}}{2}
  \end{array}
\right)+\lambda \left(
  \begin{array}{ccc}
    2-\sqrt{2}                        & -(2\sqrt{2}-2)                    & 0                    \\
    2-\sqrt{2}-\frac{1}{\sqrt{6}}A    & \sqrt{2}-1+\frac{1}{\sqrt{3}}A    & -\frac{\sqrt{2}}{2}A \\
    -(2-\sqrt{2})-\frac{1}{\sqrt{6}}A & -(\sqrt{2}-1)+\frac{1}{\sqrt{3}}A & \frac{\sqrt{2}}{2}A
  \end{array}
\right)\nonumber\\ &&+\mathcal{O}(\lambda^2).\label{ue2}
\end{eqnarray}

\begin{eqnarray}
U&=&\left(
  \begin{array}{ccc}
    \frac{\sqrt{6}}{3}  & \frac{\sqrt{3}}{3}  & 0                  \\
    -\frac{\sqrt{6}}{6} & \frac{\sqrt{3}}{3}  & \frac{\sqrt{2}}{2} \\
    \frac{\sqrt{6}}{6}  & -\frac{\sqrt{3}}{3} & \frac{\sqrt{2}}{2}
  \end{array}
\right)\nonumber\\&& +\lambda \left(
  \begin{array}{ccc}
    2-\sqrt{2} & -(2\sqrt{2}-2) & z'^\ast A \\
    2-\sqrt{2}-\left(\frac{1}{\sqrt{6}}+\frac{z'}{\sqrt{3}}\right)A
    & \sqrt{2}-1+\left(\frac{1}{\sqrt{3}}-\frac{z'}{\sqrt{6}}\right)A    & -\frac{\sqrt{2}}{2}A \\
    -(2-\sqrt{2})-\left(\frac{1}{\sqrt{6}}+\frac{z'}{\sqrt{3}}\right)A
    & -(\sqrt{2}-1)+\left(\frac{1}{\sqrt{3}}-\frac{z'}{\sqrt{6}}\right)A & \frac{\sqrt{2}}{2}A
  \end{array}
\right)\nonumber\\
&&+\mathcal{O}(\lambda^2).\label{ue1}
\end{eqnarray}
More details can be found in Ref.\cite{LISW}.

More recently, Xiao-Gang He, Shi-Wen Li and I\cite{he} applied the
new ``triminimal" parametrization method\cite{pakvasa} to both CKM
and PMNS mixing matrices. In this new method, the parameters chosen
are not the traditional deviations of the matrix elements around
unit matrix, instead, they are the deviations from big mixing angles
based on a certain mixing pattern as zeroth-order bases. The method
pointed out a new way to parameterize the mixing matrix with all
angles small, i.e. the ``triminimal" parametrization\cite{pakvasa}.
We thus arrive at a unified description between different kinds of
parametrizations for quark and lepton sectors: the standard
parametrizations, the Wolfenstein-like parametrizations, and the
triminimal parametrizations.

In above, we reported recent progress on unified parametrization of
both lepton and quark mixing matrices from standard viewpoint. Now
we report on the non-standard attempt to explain neutrino
oscillations based on the idea of Lorentz violation (LV). Neutrinos
offer a promising possibility to study Lorentz violation that may
exist at the low-energy as the remnants of Planck-scale Physics. A
number of researchers studied to explain the neutrino oscillations
by the non-standard viewpoint of Lorentz violation. Coleman and
Glashow pointed out that neutrino oscillation can take place even
for massless neutrinos if Lorentz invariance is violated in the
neutrino sector~\cite{Coleman:1997plb}. There have been more works
along this directions~\cite{Coleman:1999prd}.

Recently, Zhi Xiao, Shimin Yang and I\cite{XY} studied Lorentz
violation contribution to neutrino oscillation. In our calculation
we assume that neutrinos are massless and that the neutrino flavor
states are mixing states of energy eigenstates. We calculate
neutrino oscillation probabilities by the effective theory for
Lorentz violation, which is usually called the standard model
extension (SME)~\cite{Kostelecky:1997prd}. In our work, the mixing
angles for neutrinos are functions of Lorentz violation parameters.

Here we only report our qualitative conclusion from our study: we
carried out Lorentz violation contribution to neutrino oscillation
by the effective field theory for LV and give out the equations of
neutrino oscillation probabilities. In our model, neutrino
oscillations do not have drastic oscillation at low energy and
oscillations still exist at high energy. The oscillation amplitude
varies in different energy scale and will go to zero when the
neutrino energy is high enough. Neutrinos may have small mass and
both LV and the conventional oscillation mechanisms contribute to
neutrino oscillation. However, our calculation at the high energy
range where neutrino masses can be neglected is still applicable.

As summary, we give the following conclusions. Standard neutrino
oscillations model are well organized by PMNS matrix, which can be
unified parameterized with quark mixing matrix by combining with
quark-lepton complementarity. Neutrino oscillations can be also
obtained by Lorentz violation without neutrino mass. More detailed
analysis are needed to check whether Lorentz violation can be a
viable model for neutrino oscillations. From the standard mixing
viewpoint, the mixing part is independent of energy and neutrino
oscillations disappear at high energy. But in Lorentz violation
models, mixing part is the function of energy and the oscillation
amplitude varies with neutrino energy. Thus neutrino experiments
with energy dependence may distinguish between the conventional
massive neutrino scenario and the Lorentz violation scenario for
neutrino oscillations. Thus we suggest new experiments on neutrino
oscillations, such as the ANITA experiment\cite{ANITA}, to pay
special attention on the energy dependence of high energy neutrino
oscillations.

\section*{Acknowledgments}

I am very grateful to my collaborators, Xiao-gang He, Nan Li,
Shi-Wen Li, Zhi Xiao, Shimin Yang for the collaborated results in
this talk. I also thank Sang Pyo Kim and the organizers for their
invitation and warm hospitality. This work is partially supported by
National Natural Science Foundation of China (Nos.~10721063,
10575003, 10528510)).

\end{document}